\documentclass{article}
\usepackage{subfigure}
\usepackage{url}
\usepackage{amsfonts}
\usepackage{graphics}
\usepackage{graphicx}

\begin{document}

\title{A web-based tool to Analyze Semantic Similarity Networks.}
%
%
%
%
%

%
\author{Mario Cannataro, Pietro Hiram Guzzi, Marianna Milano, Pierangelo Veltri}

\maketitle
\begin{abstract}

In computational biology, biological entities such as genes or proteins are usually annotated with terms extracted from Gene Ontology (GO). The functional similarity among terms of an ontology is evaluated by using Semantic Similarity Measures (SSM).  More recently, the extensive application of SSMs yielded to the Semantic Similarity Networks (SSNs).  SSNs are edge-weighted graphs where the nodes are concepts (e.g. proteins) and each edge has an associated weight that represents the semantic similarity among related pairs of nodes.
The analysis of SSNs may reveal biologically meaningful knowledge. For these aims, the need for the introduction of tool able to manage and analyze SSN arises. Consequently we developed SSN-Analyzer a web based tool able to build and preprocess SSN.
As proof of concept we demonstrate that community detection algorithms applied to filtered (thresholded) networks, have better performances in terms of biological relevance of the results, with respect to the use of raw unfiltered networks.

\end{abstract}


\section{Introduction}
Biological informations about genes and proteins are stored into biological ontologies \cite{cannataro2013data} \cite{Guzzi2012} such as Gene Ontology (GO).
GO has gained a wide diffusion in bioinformatics and computational biology since it provides a structured and uniform vocabulary of terms, GO terms, useful to describe a domain of interest \cite{Pesquita2009}.
 Gene Ontology (GO) \cite{citeulike:256350} organizes a set of GO terms into three main taxonomies: Molecular function (MF), Biological Process (BP), and Cellular Component (CC).
Then, each GO Term is related to any number of gene products through the `` annotations process ''.  These annotations are aviable in public databases as, the Gene Ontology Annotation (GOA) database \cite{citeulike:256350}.
Annotations allow the comparison of entities focusing on semantics aspects through semantic similarity measures (SSMs) \cite{Pesquita2009} \cite{citeulike:10098454}. A SSM takes in input two or more terms of GO and produces as output a numeric value in the [0..1] interval representing their similarity. The use of SSMs to evaluate the functional similarity among gene products is becoming a common task and consequently the use of SSMs to analyze biological data is gaining a broad interest from researchers \cite{citeulike:1604991}  \cite{citeulike:10098454} .
Many analysis methonds based on use of SSMs are present in litterature, here,  we  focus on semantic similarity networks \cite{Pesquita2009}.
A semantic similarity network of proteins (SSN) is an edge-weighted graph, where the nodes rappresents the analyzed set of proteins, and  the set of edges, with a associated weight, represents the semantic similarity among related pairs of nodes.
These networks are constructed by computing some similarity value between genes or proteins and then linking nodes whose similarity is greater than zero. 
A possibile bias consists of the presence of meaningless edge  SSN related to high similarity value. Thus, a thresholding preprocessing can be improve the bilding of SSN. 
Many methods for networks thresholding exist, for example, methods based on global threshold, or based on local thresholds.
However, internal characteristics of SSMs \cite{guzzi2012cimento} bring to exclude the use of global thresholds, since small regions of relatively low similarities may be due to the characteristics of measures while proteins or genes have high similarity. Whereas the use of local threshold  may be influenced by the presence of local noise and in general may cause the presence of biases in different regions.
In a previous work we presented novel hybrid thresholding method employing both local and global approaches and based on spectral graph theory. The choice of the threshold is made by considering the highlighting of nearly-disconnected components. The evidence of the presence of these components is analyzed by calculating the eigenvalues of the Laplacian matrix \cite{citeulike:1318429} \cite{citeulike:1158897}. The choice of this simplification has a biological counterpart on the structure of biological networks. It has been proved in many works that these biological networks tend to have a modular structure in which hub proteins (i.e. relevant proteins) have many connections \cite{bertolazzi2013functional} \cite{ma2012biological} \cite{zhu2007getting}. Hub proteins usually connect small modules (or communities), i.e. small dense regions with few link to other regions \cite{su2010glay} in which proteins share a common function.
 We here presented a Web Based Tool for Analyzing Semantic Similarity Networks able to build and preprocess SSN. Furthermore, 
 SSN-Analyzer enables the calculation of semantic similarity matrices from a proteins/genes input dataset on which the SSNs are constructed and analyzed.
 In this way the user can easily preform the testing of interest.
 The web tool is realized with R Shiny package.



\section{Semantic Similarity Measures}
Semantic Similarity measures are instrument generally applied  to Gene ontology Terms (GO terms) in order to quantify the similarity of two or more terms of belonging ontology.
Since genes and proteins are annotated by set of GO terms, these measures are carried on genes and proteins.
In specific, the SSMs are mathematical functions able to associate a numerical value for each couple of terms of the same ontology  quantifying their similarity according to following definition:
\begin{equation}
SS_M: T x T  \rightarrow R 
 T:{t_1....t_2 }, t_i \in O
\end{equation}


In letterature there are more than 40 different semantic similarity measures and many ones rely on concepts from information theory such as information content (IC). The IC  measures how specific and informative is any term according to  the frequency of the occurrence of this term into the corpus of annotations considering the Gene Ontology Database:
\begin{equation}
IC(c) = -log(p(c))
\end{equation}
where p(c) is the probability of the occurrence of c \cite{citeulike:1238}. Thus, according to above formulation, rare term contains a greater amount of information.
The IC-based measures usually take as input two terms and then calculate the IC of a common ancestor of them.
The  ancestor can be the common ancestor with the maximum value of IC (Maximum Informative Common Ancestor MICA) or disjoint common ancestor (technical DCA). This approach  considers  all disjoint common ancestors that does not incorporate any other ancestor including the semantic tree structure of the terms. This kind of IC evaluation is also indicated as GraSM (Graph-based Similarity Measure) option.

For example Resnik meausure applied \cite{citeulike:4924299} to t1 and t2 terms considers the IC of MICA:
\begin{equation}
sim_Res(c1,c2)=IC(c_MICA)
\end{equation}

Others examples of measures based on IC of MICA are Lin measure \cite{citeulike:697707} and Relevance measure \cite{citeulike:415440}, whereas the GrasM approch can be applied to Resnik, Lin and Jiang-Conrath measures.
Also, there are semantic similarity measures based on topological properties of GO DAG.\\
The Czekanowski-Dice measure \cite{citeulike:1625362}, for example,  calculates  the distance of two terms t1 and t2 into the GO structure, and the subsequent mapping of the distance into similarity indices, i.e. the higher distance the lower similarity.
The Kappa measure \cite{citeulike:1943745} represents each gene in GO as n-dimensional vector, where each n-dimension identifies an ontology annotation. The similarity measure among two genes is calculated by taking into account the occurrence of similar annotations in the representative vectors.
 Similarly,  the Cosine similarity \cite{cho2007semantic} represents each gene g is  as a n-dimensional vector in which each component i represents the information content of the annotation i. The similarity of two genes is defined as the cosine of the angle between their vectorial representation.
Finally the Weighted-Jaccard (also known as SimGIC measure)  considers  the information content of a group of GO terms. It directly evaluates the similarity between two sets of GO terms t1 and t2 considering the contributions of all the shared ancestors of the two sets. Thus, it can be directly applied to proteins and genes.

\section{Semantic Similarity Networks}
We developed a novel hybrid thresholding method exploiting both local approch,that computes a different threshold for each node or group of nodes and global one,that prune all edges with weights lower than the threshold. Moreovere this hybrid thresholding method refers to spectral graph theory.
The algorithm examines each node of a input graph and stores the weights of adjacent edge. Then a theshold $ k= \mu + \alpha \times sd $ is deteminated, where $\alpha$ is a global variable, and  $\mu$ and sd are  the average and standard deviation of all weights, i.e local components.

Then,the algorithm compares each edge weight with threshold k: if the weight is greater than k considering the adjacent of both its nodes, it will be inserted into a novel graph with  weight 1, whereas if the weight o is greater than k considering only one of its adjacent nodes, it will be inserted into a novel graph with weight 0,5. 
When all weights of egdes are analyzed, the Laplacian of the spectrum of the graph is analyzed \cite{citeulike:1318429,citeulike:1158897} in order to evaluate the presence of nearly disconnected components. If the graph presents nearly disconnected components, the algorithm stops, alternatively a  more stringent threshold k is generated as well as a novel graph.

\subsection{Pruning Semantic Similarity Network}
In detail, the pruning algorithm examines each node \emph{i} of input graph $Gssu$ and stores all weights of the adjacent edges. After this step it determines a local threshold. Then, the algorithm inserts the node \emph{i} and all the adjacent ones in to final graph output of pruning $Gpr$.
The algorihtm  analyzes each edge adjacent to \emph{i} and inserts into $Gpr$  those with weight greater then the determined local threshold. If the considered edge is not present in $Gpr$, the edge will have weight 0,5, otherwise the weight of the edge is set to 1. Finally all the nodes with degree = 1 are deleted from $Gpr$.
In other words, if the edges  are relevant considering the neighborhood of both nodes they will compare in the pruned graph with unitary weight while if the edges  are relevant considering one node, they will compare with 0.5 weight.
For instance by setting $\alpha$ = 0 the threshold $ k=\mu+\alpha \times sd $ results equal to the average of the weights of edges adjacent to node \emph{i} $ \in  Gssu$, $AVG(node_\emph{i})$.
The algorithm initially explores each node and discarded from the analysis the ones with  degree 1. 
Then it explores the current node and their neighbors. The nodes are add in $Gpr$, and the edges in the pruned graph have a weight 0,5 or 1 according to the the average of the weights of the neighbours nodes.
In the last step all the nodes with zero degree are eliminated from $Gpr$, producing the final graph.
The generation of pruned graph is repeated until the graph has nearly disconnected components,by analyzing the spectrum of the associated laplacian for value of threshold.

\subsection{Thresholding Semantic Similarity Networks}
We developed a novel hybrid thresholding method exploiting both local approch,that computes a different threshold for each node or group of nodes and global one,that prune all edges with weights lower than the threshold. Moreovere this hybrid thresholding method refers to spectral graph theory.
The algorithm examines each node of a input graph and stores the weights of adjacent edge. Then a theshold $ k= \mu + \alpha \times sd $ is deteminated, where $\alpha$ is a global variable, and  $\mu$ and sd are  the average and standard deviation of all weights, i.e local components.

Then,the algorithm compares each edge weight with threshold k: if the weight is greater than k considering the adjacent of both its nodes, it will be inserted into a novel graph with  weight 1, whereas if the weight o is greater than k considering only one of its adjacent nodes, it will be inserted into a novel graph with weight 0,5. 
When all weights of egdes are analyzed, the Laplacian of the spectrum of the graph is analyzed \cite{citeulike:1318429,citeulike:1158897} in order to evaluate the presence of nearly disconnected components. If the graph presents nearly disconnected components, the algorithm stops, alternatively a  more stringent threshold k is generated as well as a novel graph.

\subsection{Pruning Semantic Similarity Network}
In detail, the pruning algorithm examines each node \emph{i} of input graph $Gssu$ and stores all weights of the adjacent edges. After this step it determines a local threshold. Then, the algorihtm inserts the node \emph{i} and all the adjacent ones in to final graph output of pruning $Gpr$.
The algorihtm  analyzes each edge adjacent to \emph{i} and inserts into $Gpr$  those with weight greater then the determined local threshold. If the considered edge is not present in $Gpr$, the edge will have weight 0,5, otherwise the weight of the edge is set to 1. Finally all the nodes with degree = 1 are deleted from $Gpr$.
In other words, if the edges  are relevant considering the neighborhood of both nodes they will compare in the pruned graph with unitary weight while if the edges  are relevant considering one node, they will compare with 0.5 weight.
For instance by setting $\alpha$ = 0 the threshold $ k=\mu+\alpha \times sd $ results equal to the average of the weights of edges adjacent to node \emph{i} $ \in  Gssu$, $AVG(node_\emph{i})$.
The algorithm initially explores each node and discarded from the analysis the ones with  degree 1. 
Then it explores the current node and their neighbors. The nodes are add in $Gpr$, and the edges in the pruned graph have a weight 0,5 or 1 according to the the average of the weights of the neighbours nodes.
In the last step all the nodes with zero degree are eliminated from $Gpr$, producing the final graph.
The generation of pruned graph is repeated until the graph has nearly disconnected components,by analyzing the spectrum of the associated laplacian for value of threshold.


\section{Architecture and Implementation of SSN-Analyzer}

SSN-Analyzer, a Web Based Tool for Managing Semantic Similarity Networks for bilding and preprocessing Semantic Similarity Networks

SSN-Analyzer has been implemented using the Shiny framework and the R statistical language. The tool enables the calculation of semantic similarity form input genes/proteins dataset as well as the construction of SSN and preprocessing step.The tool provides a simple Graphical User Interface allowing the user an easy access to the tool functionalities.

\begin{figure}[ht]
\centering
 {\includegraphics[width=1.5in]{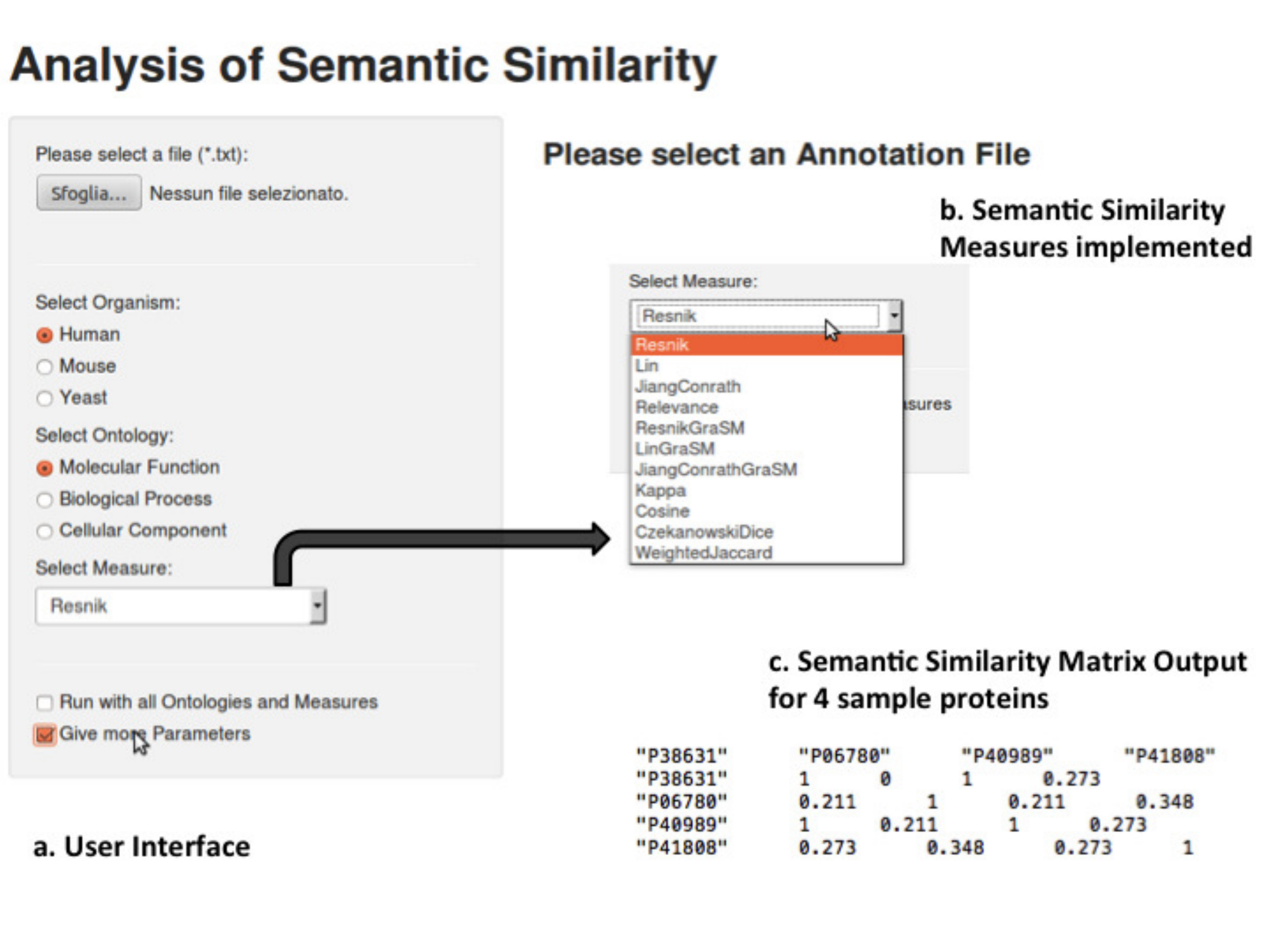}}
   \caption{Semantic Similarity Analysis Interface}
\label{fig:step1}
\end{figure}

Figure \ref{fig:step1} conveys the interface where the user can execute a semantic similarity analysis.
The tool requires as input data a list of proteins and the related annotations for each ones. It is possibile to select the organism of genes/proteins dataset, the ontology MF, BP, CC and a semantic similarity measure. Otherwise, once the organism is selected, there is the option to run the analysis by using all available semantic similary measures on each ontology.  

About the SSMs, the tool provides the measures implemented in  R package csbl.go such as Resnik,ResnikGraSM, Lin, Lin-with the GraSM option (LinGraSM here after), JiangConrath, JiangConrath with the GraSM option (JiangConrathGraSM here after), Relevance, Kappa, Cosine, WeightedJaccard, and Czekanowski Dice.

\begin{figure}[ht]
\centering
{\includegraphics[width=3.5in]{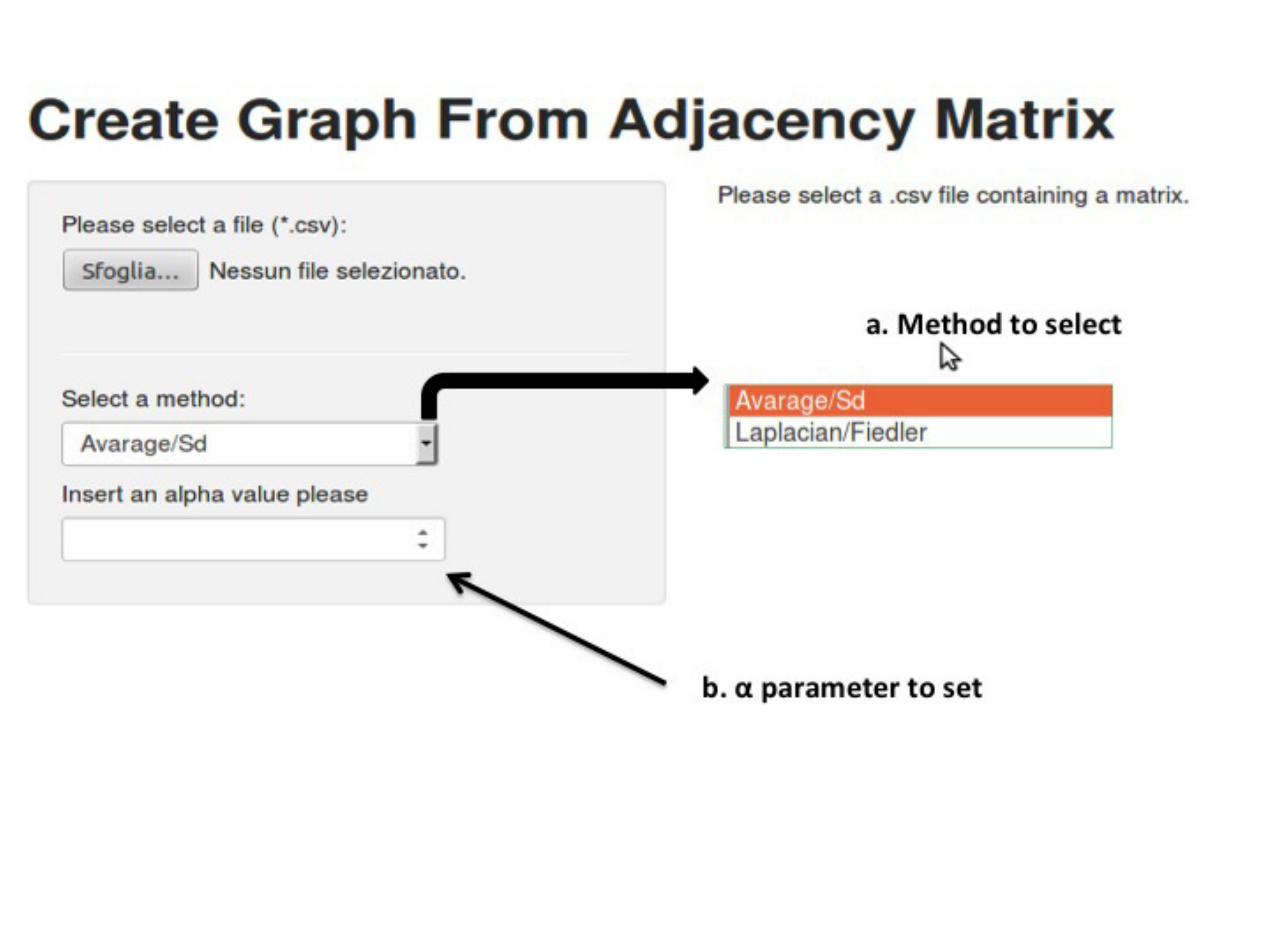}}
   \caption{ Graph Generation Interface}
\label{fig:step3}
\end{figure}
\begin{figure*}[ht]
\centering
 {\includegraphics[width=4.5in]{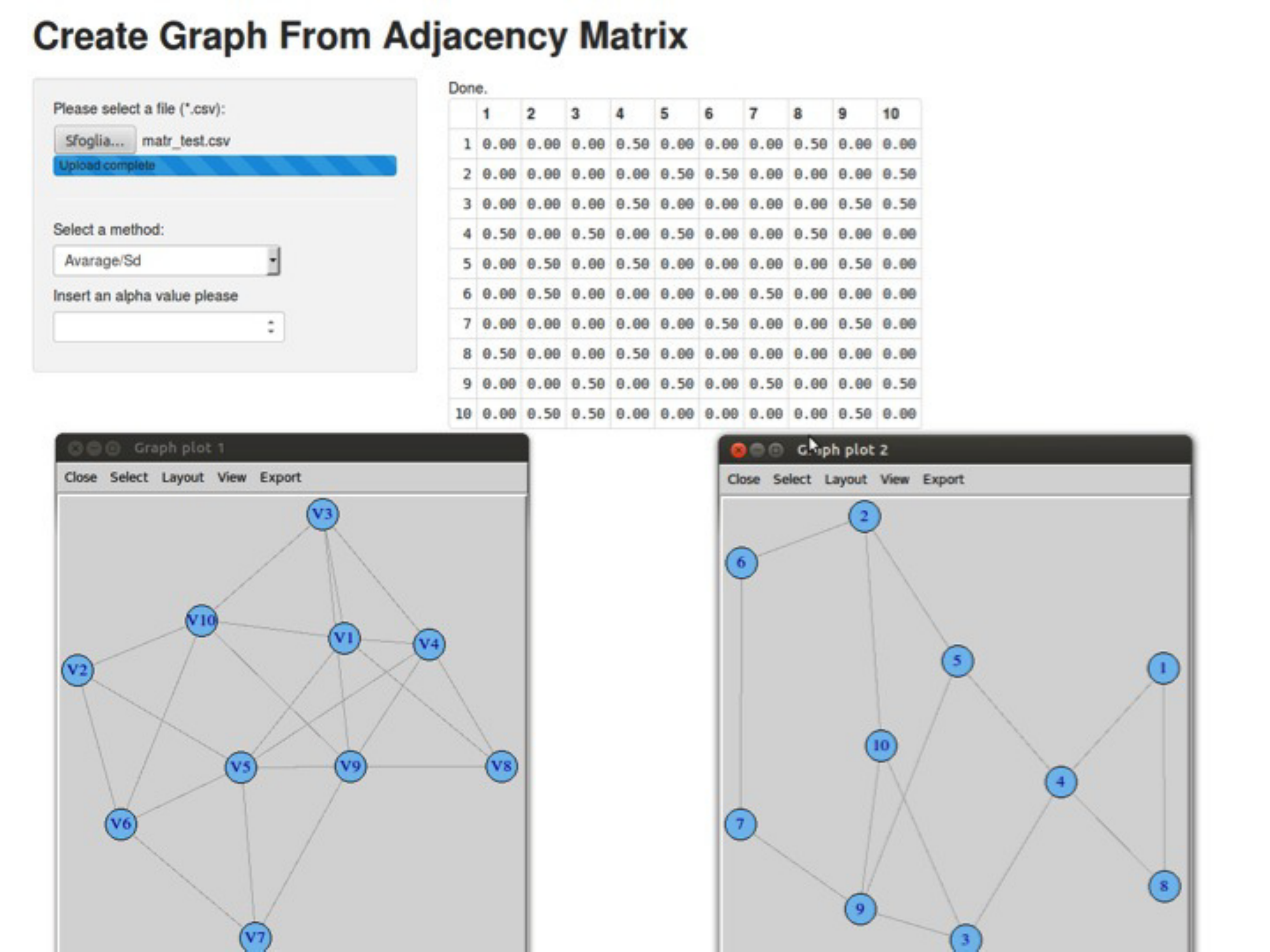}}
   \caption{ Thresholded Matrix. The first graph is related to no-thresholded initial matrix, the second one is resulted graph after pruning algorithm}
\label{fig:step4}
\end{figure*}

The output file (\ref{fig:step1})(c) is a semantic similarity matrix use to built the SSN.
Figure \ref{fig:step3} shows the interface where the user inserts a semantic similarity network in order to perform preprocessing process.
 Figure \ref{fig:step4} conveys the output matrix and the output graph: the thresholded adjacency matrix  is the result of the preprocessing analysis and the related pruned graph. Moreover, there is the option for the visualization of raw graph in order to assess the results for future analysis.

\subsection{Architecture}

The Shiny is a R library able to extend the R functionality allowing the conversion of  a R code  into an interactive website user friendly. In shiny, the user can create a Graphic User Interface and publish it on web page through the extension Shiny-server. The Shiny web  interfaces are automatically loaded in live mode, thus, the users can be insert any additional input parameter and it is not necessary to reload the browser page.
Furthermore, several libraries are integrate very intuitive widgets optimized for Shiny, in this way the user can make  simple and functional  graphical  interface.
An R Shiny project consists of two scripts, ui.R (user-interface) and server.R (server-side).

 Server.R script contains the R code, while ui.R Script defines the GUI that is displayed in the brower. 
In ui.R headerPanel (), sidebarPanel (), Mainpanel () are definable, and the user can build the inputs and the output.
When the server.R is created it is necessary to define the logic of the server and make functional the various input and output stated in ui.R. In server.R the user  imports the necessary libraries, creates the  functions and imports the inputs. Then, the user assigns the  outputs in order to link the results of the file server and view them in the GUI. 
Several libraries ad hoc can be integrated allowing the specific analysis according to own data.
An example of Shiny project for a Web Based Tool for Thresholding Semantic Similarity Networks, which we developed, is reported below:

\begin{figure}[ht]
\centering
{\includegraphics[width=2.5in]{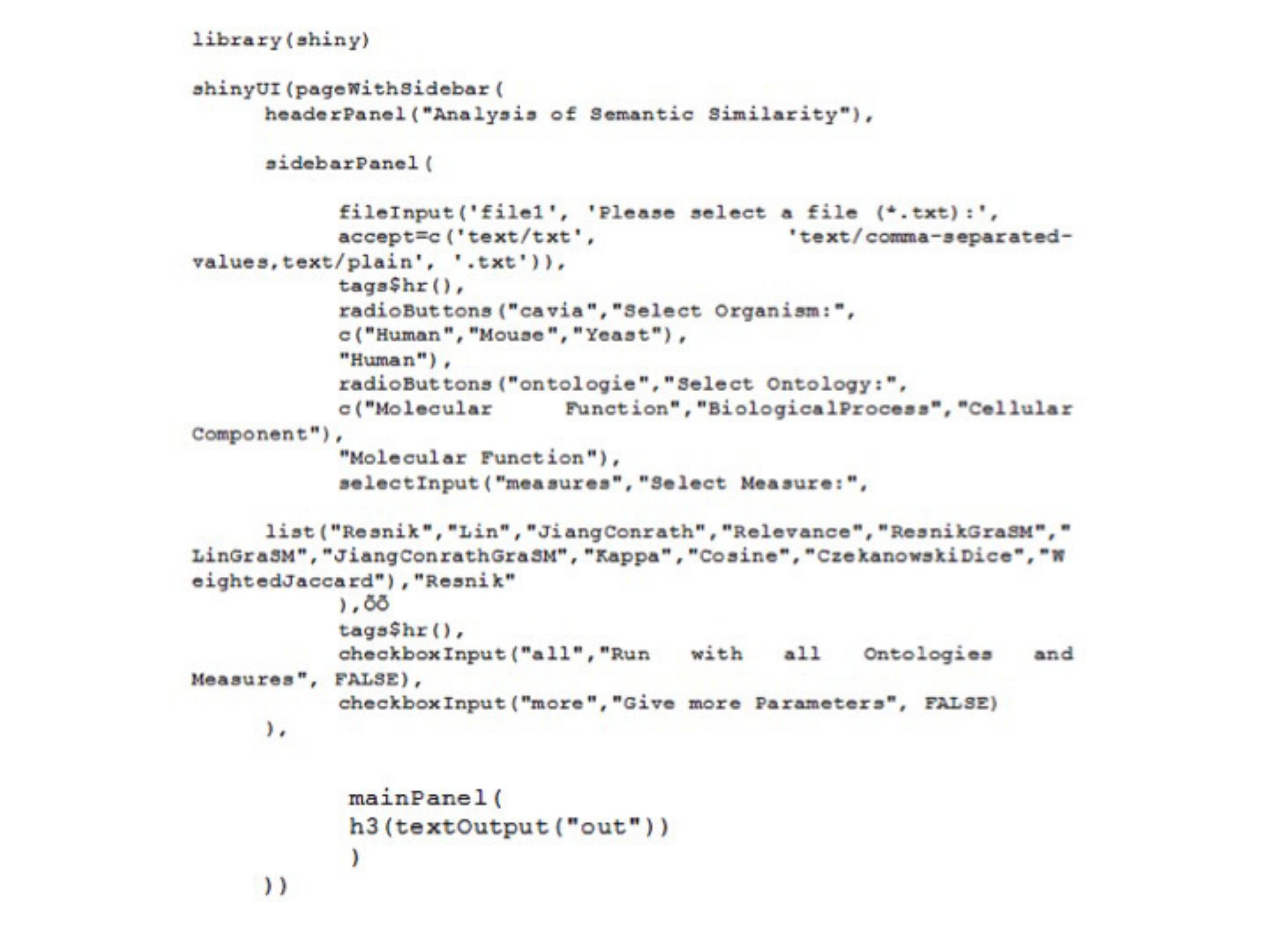}}
   \caption{ ui.R Script Example}
\label{fig:ui}
\end{figure}

\begin{figure*}[ht]
\centering
{\includegraphics[width=5in]{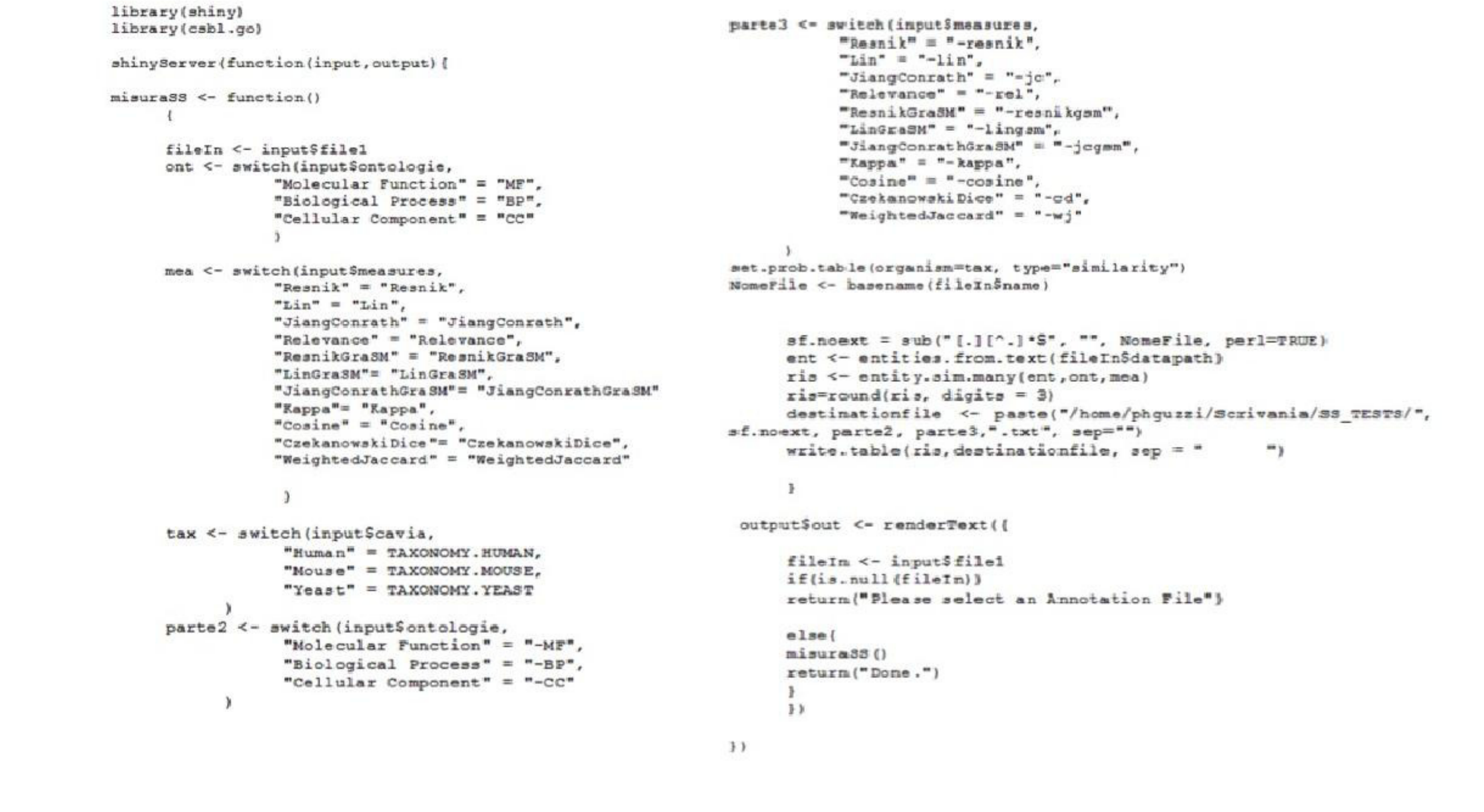}}
   \caption{server.R Script Example}
\label{fig:server}
\end{figure*}

\section{Conclusions}
Gene ontology and annotations are widely used in bioinformatics especially in the quantitative comparison of genes and proteins functions leveraging on the measures of semantic similarity between two or more functional genes/proteins, defined by the GO terms associated with them.
Semantic Similarity Networks of proteins are a powerful instrument for biological research, expecially in analysis of organisms on a system scale.
Due to semantic similarity values, that weigh the relations among proteins, the SSNs  may contain meaningless edges; a preprocessing step may be necessary in order to improve the SSNs performance. For example, thresholding algorithms prune the weighted edge of graph according to a set threshold.
thus they may be pruned using thresholding algorithms.
In this paper we presented SSN-Analyzer, a Web Based Tool for Managing Semantic Similarity Networks able to perform  a semantic similarity analysis on proteins/genes of specific organisms and build SSNs.  Furthermore the tool is able to apply a novel thresholding technique, that we developed, on SSNs in order to achieve more meaningful information.
The tool can be help the users to perform easily semantic similarity testing and improve the results for future analysis on SSNs.

\section*{Acknowledgments}

This work has been partially founded by project PON Smartcities DICET-INMOTO-ORCHESTRA PON04a2\_D.


%

%
%





\end{document}